\documentclass[conference]{IEEEtran}
\IEEEoverridecommandlockouts
\usepackage{algorithm}
\usepackage{algpseudocode}
\usepackage{cite}
\usepackage{amsmath,amssymb,amsfonts}
\usepackage{graphicx}
\usepackage{textcomp}
\usepackage{xcolor}
\def\BibTeX{{\rm B\kern-.05em{\sc i\kern-.025em b}\kern-.08em
    T\kern-.1667em\lower.7ex\hbox{E}\kern-.125emX}}
    
\begin{document}
\fontsize{9pt}{10.5pt}\selectfont
\title{ASAP: An Azimuth-Priority Strip-Based Search Approach to Planar Microphone Array DOA Estimation in 3D\\

\thanks{Ming Huang and Shuting Xu contribute equally to this work. This work was supported by the National Key R$\&$D Program of China under Grant No. 2024YFB4710900, the National Natural Science Foundation of China under Grant No. U24A20265, the Open Research Fund of the State Key Laboratory of Multimodal Artificial Intelligence Systems, the Shenzhen Science and Technology Program under Grant No. KQTD20221101093557010, and the Guangdong Science and Technology Program under Grant No.2024B1212010002. This research was conducted when Shuting Xu, Leying Yang, and Huanzhang Hu were visiting students at the Southern University of Science and Technology, Shenzhen, China. Corresponding author: He Kong (email: kongh@sustech.edu.cn).
}
}

\author{
    \IEEEauthorblockN{Ming Huang\textsuperscript{1}, Shuting Xu\textsuperscript{1}, Leying Yang\textsuperscript{1}, Huanzhang Hu\textsuperscript{1}, Yujie Zhang\textsuperscript{1}, Jiang Wang\textsuperscript{2}, 
    Yu Liu\textsuperscript{1}\\
    Hao Zhao\textsuperscript{1}, and He Kong\textsuperscript{1}}

\IEEEauthorblockA{\textsuperscript{1}\textit{Southern University of Science and Technology, Shenzhen, China}}
\IEEEauthorblockA{\textsuperscript{2}\textit{Institute of Science Tokyo, Japan}}
}

\maketitle

\begin{abstract}
Direction-of-arrival (DOA) estimation is an important task in microphone array processing and many downstream applications. The steered response power with phase transform (SRP-PHAT) method has been widely adopted for DOA estimation in recent years. 
However, accurate SRP-PHAT estimation in 3D scenarios requires evaluating steering responses over thousands of candidate directions, severely limiting real-time performance on resource-constrained platforms. This challenge becomes even more critical for planar arrays, which are widely used in robotics due to their structural simplicity. Motivated by the fact that azimuth estimation is usually more reliable than elevation estimation for most arrays, we propose ASAP, an \underline{A}zimuth-priority \underline{S}trip-based Se\underline{a}rch a\underline{p}proach to planar microphone array DOA estimation in 3D. In the first stage, ASAP performs coarse-to-fine region contraction within azimuthal strips to lock azimuth angles while retaining multiple maxima through spherical caps. In the second stage, it refines elevation along the great-circle arc between two close candidates. Extensive simulations and real-world experiments validate the efficiency and merits of the proposed method over existing approaches.
\end{abstract}


\section{Introduction}
\label{sec:intro}
Direction-of-arrival (DOA) estimation for microphone arrays enables acoustic perception in many applications, including  teleconferencing \cite{Itzhak2025TASLP,Gburrek2022IWAENC,Yoshioka2019ADM,TASLP2025}, target tracking \cite{Li2012FADE,Ho2007Accurate,grumiaux2022survey,Rakhshan2025Observability}, environmental monitoring \cite{wang2026manifold,li2025acousticcamera,fu24,Zhao2023JEA}, robot audition \cite{Jiang25,TDOA2025,Liu2025SAGENet}, and human–robot interaction \cite{su2021necessary,zhang2025hybridtdoa,Kulic2025}. Among the existing approaches, the steered-response power with phase transform (SRP-PHAT) method remains a strong baseline \cite{DiBiase2000Thesis}. SRP-PHAT extends the generalized cross-correlation with phase transform (GCC-PHAT) algorithm by normalizing spectral magnitudes, ensuring equal frequency weighting and sharper correlation peaks for DOA estimation \cite{Dmochowski2007GSRP,Knapp1976GCC}. SRP-PHAT tolerates reverberation and noise with moderate modeling assumptions \cite{Dang2024IRSRP,Wang2024}, and can be easily integrated into more advanced frameworks \cite{Grinstein2024_EURASIP_SRP_Tutorial,DiazGuerra2018SRPNN, Pang2025Hybrid,Wang2014NLOS,Wang2023Elliptic}. 


Although SRP-PHAT is recognized for its robustness, it is computationally demanding, particularly in 3D scenarios. In particular, SRP-PHAT performs exhaustive evaluation over dense spherical grids, often comprising thousands of candidate directions, which can be prohibitive for embedded or mobile platforms with limited computing resources \cite{Jiang25,fu24,Liu2025SAGENet,zotkin2004,wang2025singlemic}. Moreover, planar arrays are preferred for their structural simplicity and ease of integration with the robotic hardware system \cite{Li2003Design,Zhao2023JEA}. For planar arrays, elevation estimates are usually less reliable than the azimuth estimates, leading to ambiguities and
reduced uniformity in 3D \cite{Moore2024Auditory,Fu2025AuralNet}.


Despite the wide applicability of planar microphone arrays, developing an efficient approach to 3D DOA estimation remains a critical challenge. To reduce the computational burden, prior studies have proposed two complementary families of methods. On the one hand, volumetric SRP and its variants evaluate regions rather than individual points, enabling coarser initial sampling without sacrificing robustness; a refinement step then recursively subdivides high-likelihood cells and re-evaluates them \cite{Cobos2011ModifiedSRPPHAT,lima2015,Nunes2014HSRP}. 

On the other hand, region contraction methods progressively shrink the candidate set around high-scoring maxima. Relevant methods include stochastic region contraction (SRC) \cite{Do2007src} and its deterministic counterpart, coarse-to-fine region contraction (CFRC) \cite{Do2007cfrc}. Region contraction methods iteratively rescore only a small subset at increasing resolution, thus achieving substantial computational savings while maintaining localization accuracy. 

However, directly applying existing methods to planar arrays might be challenging. For example, full-grid SRP-PHAT is computationally prohibitive for real-time use. while recent methods such as CFRC treat the 3D space uniformly—failing to exploit the planar array's better azimuth resolution than elevation resolution. To address these limitations, we propose ASAP, an \underline{A}zimuth-priority \underline{S}trip-based Se\underline{a}rch A\underline{p}proach to planar microphone array DOA estimation in 3D.


Rather than exhaustively evaluating thousands of directions as in full-grid SRP-PHAT, or relying on unconstrained multi-stage heuristics such as CFRC \cite{Do2007cfrc}, we decompose the 3D search into two stages. In the first stage, ASAP performs a CFRC-style scan within azimuthal strips and applies spherical-cap filtering to concentrate evaluations near emerging peaks. The second stage then estimates elevation through a one-dimensional pass around the locked azimuth. 

Unlike existing sequential 2D methods \cite{Zhang2021Reduced}, ASAP integrates region contraction into the azimuth stage via strip scanning and spherical-cap filtering; by explicitly exploiting the planar array's anisotropic reliability, it introduces the structured guidance necessary to effectively avoid full 3D scans.


Extensive simulations and real-world experiments validate ASAP’s efficiency and performance improvement against SRP-PHAT and CFRC \cite{Do2007cfrc}. In simulations at the highest grid resolution, ASAP reduces runtime by 13.56\% and RMSE by 5.87\%, compared to CFRC \cite{Do2007cfrc}. Real-world tests with an 8-microphone uniform circular array (UCA) further show 13.98\% faster runtime and 4.33\% lower RMSE than CFRC \cite{Do2007cfrc}. Moreover, ASAP consistently achieves lower angular errors with faster runtime than full-grid SRP-PHAT across various resolution settings. To benefit the community, we open-source all the codes at https://github.com/AISLAB-sustech/ASAP/tree/main.

\vspace{-4mm}

\begin{figure}[htbp]
    \centering
\includegraphics[width=1.01\columnwidth, height=0.26\columnwidth]{Figure_1_re.png}
    \vspace{-8mm}  
    \caption{Overall architecture of the proposed ASAP framework. Stage 1 estimates a coarse azimuth $\hat{\phi}$ through strip-constrained CFRC search. Stage 2 refines the estimates from Stage 1.}
    \label{fig:pipeline}
\end{figure}

\section{THE PROPOSED METHOD}

\subsection{Problem Formulation}
We consider a UCA consisting of $M$ microphones arranged on a horizontal plane with radius $R$ \cite{Li2003Design}. The position of the \(m\)-th microphone, for \(m=1,2,\ldots, M\), can be expressed as
\begin{equation}  \mathbf{r}_m = [R\cos\theta_m,\; R\sin\theta_m,\; 0]^T,\quad \theta_m = \tfrac{2\pi(m-1)}{M}. \end{equation} 

For a sound source with azimuth angle \(\phi\in [-180^\circ,180^\circ)\) and elevation angle \(\theta\in [0^\circ,90^\circ]\), we denote the received signal by the \(m\)-th microphone at time \(t\) as $x_m(t,\theta,\phi)$. As in \cite{DiBiase2000Thesis,
Dmochowski2007GSRP}, we use $P\left(\phi,\theta\right)$ to represent the steered response power with the use of the phase transform, namely,
\begin{equation}  P\left(\phi,\theta\right)=2\pi\sum_{l=1}^{M-1}\sum_{m=l+1}^{M}R_{lm}\left(\Delta\tau_{lm}\left(\mathbf{u}\left(\phi,\theta\right)\right)\right),  \end{equation}
where $R_{lm}$ is the GCC-PHAT between $x_l(t,\theta,\phi)$ and $x_m(t,\theta,\phi)$ received by the $l$-th and the $m$-th microphone, respectively; $\mathbf{u}$ is the unit direction vector and takes the form of \begin{equation} \mathbf{u}(\phi,\theta)= \begin{bmatrix} \cos\phi\cos\theta\\ \sin\phi\cos\theta\\ \sin\theta \end{bmatrix}, \end{equation} $\Delta\tau_{lm}\left(\mathbf{u}\right)=\tau_{l}\left(\mathbf{u}\right)-\tau_{m}\left(\mathbf{u}\right)$, in which $\tau_{l}\left(\mathbf{u}\right)$ and $\tau_{m}\left(\mathbf{u}\right)$ are the time delays occurring from the direction $\mathbf{u}$ to the $l$-th and $m$-th microphones, respectively. Given the above system model, our primary objective is to efficiently and accurately estimate the azimuth and elevation angles $\hat{\phi} \text{ and }\hat{\theta}$ that maximize the steered response power $P\left(\phi,\theta\right)$
as follows: 
\begin{equation}
\hat{\phi},\hat{\theta} = \arg\max_{\substack{\phi , \theta}} P\left(\phi,\theta\right).
\label{eq:doa_estimation}
\end{equation}
\subsection{Azimuth-Prioritized Strip-Based Search}
A limitation, long noted in array signal processing and acoustic source localization, is that the mainlobe of an SRP-PHAT beamformer broadens and peaks flatten at higher elevation angles, resulting in reduced elevation accuracy compared to azimuth \cite{DiBiase2000Thesis}. Based on this observation, we propose ASAP (shown in Fig.~\ref{fig:pipeline}), which comprises two stages: a CFRC-style coarse azimuth over azimuthal strips is performed, followed by 1D elevation refinement, reducing complexity without compromising robustness.

\paragraph{\textbf{Stage 1: Strip-Constrained Azimuth Angle Estimation}} The candidate space is restricted to a union of horizontal strips centered at a predefined set. Let \(\Theta=\{\theta^{(k)}\}_{k=1}^{K}\subset[0^\circ,90^\circ]\) be the set of strip centers, with \(K:=|\Theta|\). 
Denote \(\Delta\theta>0\) as the half-width and   
\begin{equation} \Omega_{0}=\bigcup_{k=1}^{K}\big\{(\phi,\theta):\,|\theta-\theta^{(k)}|\le\Delta\theta\big\}. \end{equation} 

Within $\Omega_0$, a CFRC-style~\cite{Do2007cfrc} search is executed. More specifically, at refinement level $i=1,\ldots,L$ (with $L$ the total number of levels), we retain the top-$N$ unit directions $\{\mathbf u_j^{(i)}\}_{j=1}^N$ at each level. At each level $i$, SRP-PHAT scores are evaluated, the top-$N$ directions are retained, and spherical caps \begin{equation} C(\mathbf{u}_j^{(i)},\alpha_i)= \{\mathbf{u}:\arccos(\mathbf{u}\cdot \mathbf{u}_j^{(i)})\le\alpha_i\} \end{equation} are formed around these maxima. Here, \(i\in\{1,\ldots,L\}\) indexes the refinement level; \(\mathbf{u}_j^{(i)}\) denotes the \(j\)-th retained \emph{unit} direction at level \(i\) with \(j=1,\ldots,N\); \(\alpha_i>0\) is the geodesic half-angle (in radians); and \(C(\mathbf{u},\alpha)\) denotes a spherical cap, i.e., the set of unit vectors \(\mathbf v\) satisfying \(\arccos(\mathbf v\cdot \mathbf u)\le\alpha\). We denote the tessellation subdivision level by $\ell$ ($\ell=1,\ldots,L$), which determines the sampling density on the unit sphere $\mathbb{S}^2$. 

Starting from an initial spherical mesh ($\ell=1$),
which comprises uniformly distributed triangular faces, each increment in $\ell$ recursively subdivides every triangle into four smaller ones. This process scales the number of candidate points by approximately a factor of four. For example, Level 1 yields 42 points, whereas Level 5 reaches 10,242 points, providing high-resolution sampling for precise estimation. The output of Stage 1 is a coarse azimuth estimate $\hat{\phi}$, derived from the primary peak direction.

\begin{figure}[t]
    \centering
    \includegraphics[width=0.75\linewidth]{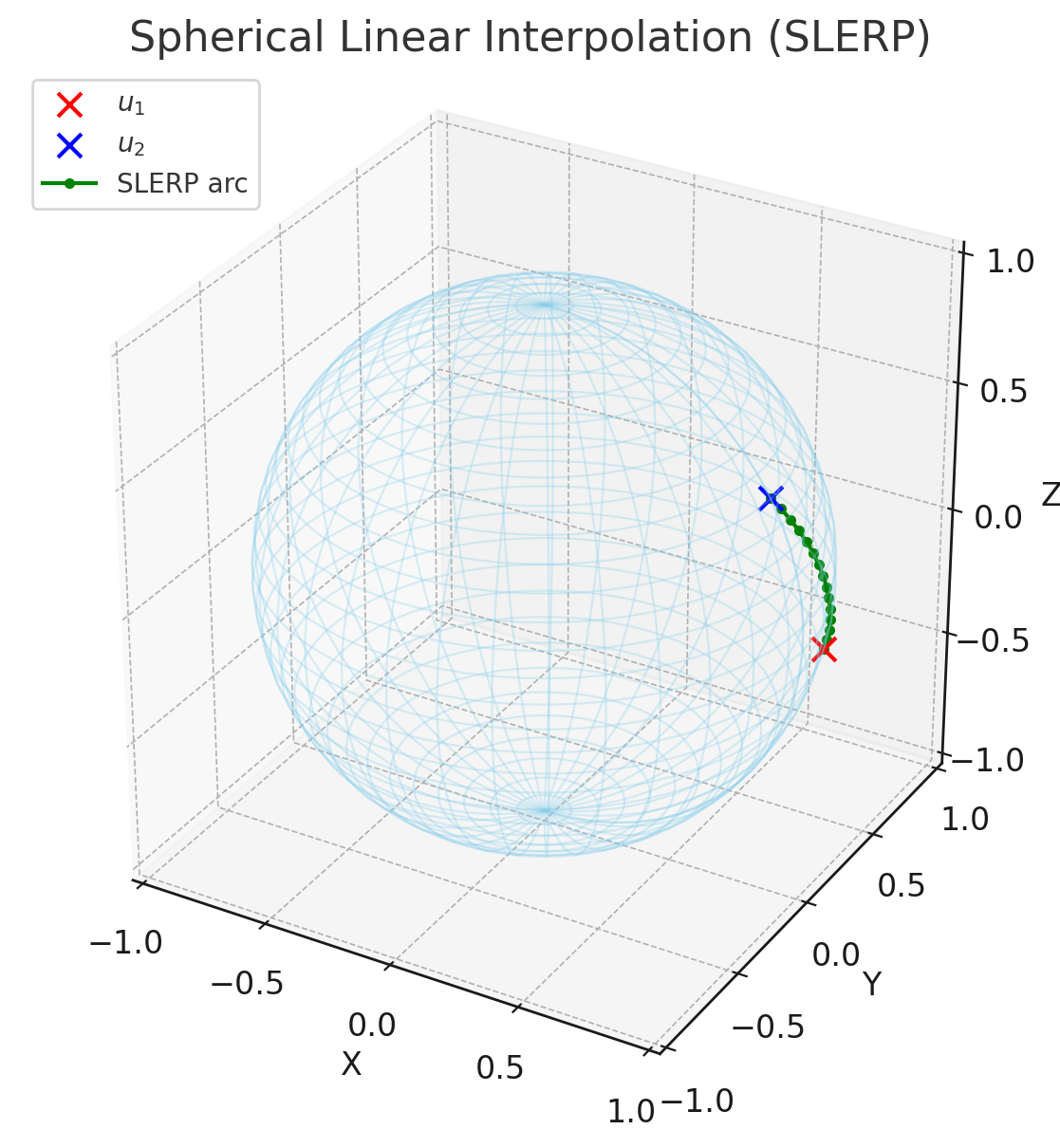}
    \vspace{-4mm}
    \caption{Illustration of spherical linear interpolation (SLERP) between two unit vectors 
    $\mathbf u_1$ (red) and $\mathbf u_2$ (blue). The green arc shows interpolated points lying on the great-circle path.}
    \vspace{-4mm}  
    \label{fig:slerp}
\end{figure}

\paragraph{\textbf{Stage 2: One-Dimensional Elevation Angle Refinement}}
Stage 2 refines elevation along a 1D path, avoiding a full 2D search and leaving the SRP-PHAT objective unchanged. Using the Stage~1 azimuth $\hat{\phi}$ and a provisional elevation $\hat{\theta}$, we restrict $\theta$ to a narrow window
\begin{equation}
\theta \in [\hat{\theta}-\Delta\theta_{2},\ \hat{\theta}+\Delta\theta_{2}].
\end{equation}

Depending on the subsequent refinement strategy, we retain either the single best unit vector \(\mathbf{u}_1\) or the top-2 unit vectors \(\mathbf{u}_1, \mathbf{u}_2\).
We propose two alternatives for refinement. 

\begin{enumerate}
\item \textbf{Meridian-Centered (MC) Refinement:} In this strategy, the azimuth is fixed in the Stage-1 estimate $\hat{\phi}$, 
and the elevation angle is searched within a bounded interval $[\theta_{\min},\theta_{\max}]$: 
\begin{equation}
  \Omega_{1} = \{(\phi,\theta) \mid \phi=\hat{\phi},\; \theta_{\min}\leq \theta \leq \theta_{\max}\}. 
\end{equation}

Here $\theta_{\min}=\max(0^\circ,\hat{\theta}-r)$ and $\theta_{\max}=\min(90^\circ,\hat{\theta}+r)$ with $r>0$ being the half-window (deg), and $h_\theta>0$ being the elevation sampling step (deg).

Candidate elevations are uniformly sampled with step size $h_\theta$, 
and the SRP-PHAT functional $P(\hat{\phi}, \theta)$ is evaluated at each grid point. 
The discrete maximizer is defined as
\begin{equation}
  \theta^\dagger = \arg\max_{\theta \in \Omega_1} P(\hat{\phi}, \theta).
\end{equation}

To improve angular resolution beyond the sampling step, a local quadratic interpolation 
is applied around $\theta^\dagger$. Let $P(\theta^\dagger)$ be the peak value and let
 $P(\theta_-), P(\theta_+)$ represent the neighboring scores at 
$\theta^\dagger - h_\theta$ and $\theta^\dagger + h_\theta$, respectively. 
The estimate of the refined elevation angle is given by
\begin{equation}
  \theta^\ast = \theta^\dagger + \frac{h_\theta}{2}\,
  \frac{P(\theta_{-})-P(\theta_{+})}{P(\theta_{-}) - 2P(\theta^\dagger) + P(\theta_{+})}. 
\end{equation}
{
\fontsize{9}{10.8}\selectfont 
We denote the neighbor samples by $\theta_\pm=\theta^\dagger\pm h_\theta$.
The final DOA estimate is $(\phi^\ast,\theta^\ast)=(\hat{\phi}, \theta^\ast)$. The MC refinement procedure is summarized in Algorithm \ref{mc}.
}
\item \textbf{Between-Points (BP) Refinement}: If Stage-1 yields two dominant unit vectors $\mathbf{u}_1,\mathbf{u}_2$, spherical linear interpolation (SLERP) generates candidates along the great-circle arc:
\begin{equation*}
  \mathbf{u}(t)=\frac{\sin((1-t)\alpha)}{\sin\alpha}\mathbf{u}_1+
  \frac{\sin(t\alpha)}{\sin\alpha}\mathbf{u}_2,\quad
  \alpha=\arccos(\mathbf{u}_1^\top\mathbf{u}_2).
\end{equation*}
In the above equation, the interpolation parameter is $t\in[0,1]$. We sample the arc with angular step $h>0$ (deg). Let $j^\dagger$ denote the index of the best-scoring arc sample.
The BP refinement procedure is detailed in Algorithm \ref{bp}.
\end{enumerate}

\begin{algorithm}[t]
\caption{Stage-2 Meridian-Centered Refinement}
\label{mc}
\small 
\begin{algorithmic}[1]
\Require Stage-1 estimate $(\hat{\phi},\hat{\theta})$, window $r$, step $h_\theta$
\Ensure Refined DOA $(\phi^\ast,\theta^\ast)$
\State $\theta_{\min}\gets \max(0,\hat{\theta}-r)$;\quad $\theta_{\max}\gets \min(90,\hat{\theta}+r)$
\State Generate grid $\{\theta_i\}$ on $[\theta_{\min},\theta_{\max}]$ with step $h_\theta$
\For{each $\theta_i$} 
  \State $P_i \gets P(\hat{\phi},\theta_i)$ 
\EndFor
\State $\theta^\dagger \gets \arg\max_{\theta_i} P_i$
\If{quadratic refinement enabled} 
  \State $\theta^\ast \gets$ refine using Eq.~(10)
\Else 
  \State $\theta^\ast \gets \theta^\dagger$
\EndIf
\State $\phi^\ast \gets \hat{\phi}$;\quad \Return $(\phi^\ast,\theta^\ast)$
\end{algorithmic}
\end{algorithm}

\begin{algorithm}[htbp]
\caption{Stage-2 Between-Points Refinement}
\label{bp}
\small 
\begin{algorithmic}[1]
\Require Stage-1 top-2 unit vectors $\mathbf{u}_1,\mathbf{u}_2$, step $h$
\Ensure Refined DOA $(\phi^\ast,\theta^\ast)$
\State Generate arc samples $\{\mathbf{u}(t_j)\}$ via SLERP with spacing $\approx h$
\For{each $\mathbf{u}(t_j)$} \State Convert to $(\phi_j,\theta_j)$ and compute $P_j$ \EndFor
\State $j^\dagger\gets\arg\max_j P_j$;\quad $(\phi^\dagger,\theta^\dagger)\gets(\phi_{j^\dagger},\theta_{j^\dagger})$
\If{quad-refine enabled} \State $(\phi^\ast,\theta^\ast)\gets\textsc{ParabolaRefineOnArc}(j^\dagger)$ \Else \State $(\phi^\ast,\theta^\ast)\gets(\phi^\dagger,\theta^\dagger)$ \EndIf
\State \Return $(\phi^\ast,\theta^\ast)$
\end{algorithmic}
\end{algorithm}

\section{SIMULATION EXPERIMENTS}
\subsection{Simulation Setup}
In our simulation setup, an UCA ($M=8$, $R=44.4$~mm) was placed on the horizontal plane. The sound source was a 0.5-s linear chirp (500--2500~Hz) sampled at $f_s=50$~kHz. Each channel received the propagated signal plus independent random background noise $n_m(t)$ per microphone. Three conditions were considered: (i) a pure Linear Frequency Modulation (LFM) signal without noise; (ii) an LFM signal corrupted by additive noise with SNR of 3.09 dB; and (iii) an LFM signal corrupted by additive noise with SNR of 1.5 dB. 

The signals were transformed using a short-time Fourier transform (STFT) with a Hanning window ($N_{\text{FFT}}=1024$, 50\% overlap), and PHAT weighting was applied. 
True directions were drawn uniformly from $\phi\in[-180^\circ,180^\circ)$ 
and $\theta\in[0^\circ,90^\circ]$. The proposed two-stage CFRC method was evaluated against two baselines: (i) full grid SRP-PHAT and (ii) CFRC without strip constraints. Performance was assessed in terms of azimuth/elevation RMSE and runtime.

\subsection{Results and Discussions}
The evaluation of different DOA estimation methods is based on a set of accuracy and efficiency-oriented metrics: 
(i) Root mean squared error (RMSE); (ii) Computation time: indicator of real-time feasibility.

\paragraph{Simulation Results on Estimation Error}

We evaluated the ASAP under the above three SNR conditions. Performance was evaluated in terms of azimuth angle and elevation angle RMSE, 
with comparisons made against the  full grid SRP-PHAT and CFRC baselines.
\vspace{-1.5mm}

\begin{table}[htbp]
\caption{RMSE (deg) of Full Grid SRP-PHAT, CFRC, BP, and ASAP for 3751 Points in Simulation. Bold means better.}
\vspace{-6mm}
\begin{center}
\renewcommand{\arraystretch}{1.3} 
\resizebox{\columnwidth}{!}{
    \begin{tabular}{|c|c|c|c|c|} 
    \hline
    \textbf{Condition} & \textbf{SRP-PHAT \cite{Knapp1976GCC}} & \textbf{CFRC \cite{Do2007cfrc}} & \textbf{BP (ASAP)} & \textbf{MC (ASAP)} \\
    \hline
    LFM & 2.79 & 3.16 & \textbf{2.73} & 3.15 \\
    \hline
    \begin{tabular}[c]{@{}c@{}} LFM + Noise \\ (3.09 dB) \end{tabular} & 3.19 & 3.25 & \textbf{3.15} & 3.23 \\
    \hline
    \begin{tabular}[c]{@{}c@{}} LFM + Noise \\ (1.5 dB) \end{tabular} & 3.30 & 3.31 & \textbf{3.28} & 3.30 \\
    \hline
    \end{tabular}
}
\label{tab:mae_results}
\end{center}
\end{table}
\vspace{-8pt} 
\begin{figure}[H]
  \centering
  \includegraphics[width=1\columnwidth, height=0.35\columnwidth]{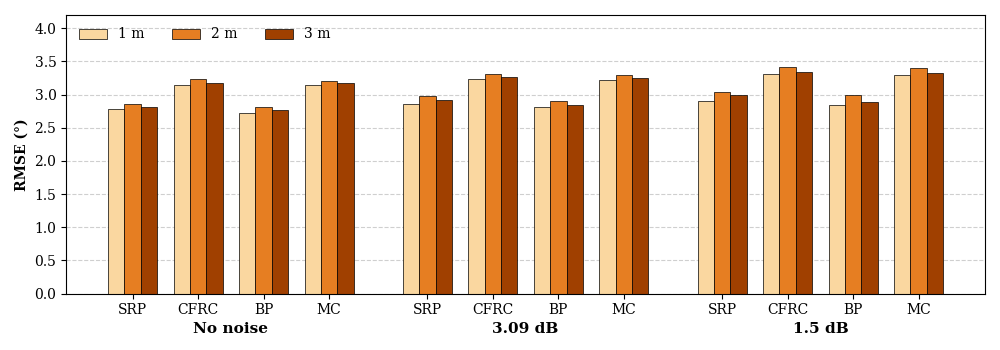}
  \vspace{-6mm}
  \caption{Comparison of RMSE under different source distances (1~m, 2~m, 3~m) 
  and noise conditions (no noise, 1.5~dB, 3.09~dB) for four algorithms: 
  full grid SRP-PHAT, CFRC, BP (ASAP), and MC (ASAP).}
  \vspace{-3mm}  
  \label{fig:mae_softcolors}
\end{figure}


Table \ref{tab:mae_results} reports the RMSE at 1~m under clean LFM signals and with additive noise at SNRs of \(1.5\)~dB and \(3.09\)~dB. Fig.~\ref{fig:mae_softcolors} further summarizes the RMSE performance under different sound source distances (1, 2, and 3 meters) and Signal-to-Noise Ratio (SNR) conditions. Among these, the proposed BP method achieves the lowest RMSE across all scenarios, outperforming SRP-PHAT and CFRC. This is because BP leverages the higher reliability of azimuth estimation for planar arrays and refines elevation via spherical linear interpolation (SLERP). The results confirm that the two-stage refinement of ASAP simultaneously enhances accuracy and robustness under varying distance and noise levels.

\paragraph{Comparison of Computational Time}
The computation time of the full grid SRP-PHAT, CFRC, BP, and MC methods was compared under different maximum refinement levels (Level 3, Level 4, and Level 5), specifically for the total computation of 3751 points in the simulation. Table \ref{tab:time_comparison} presents the quantitative results, where the best computation time for each level is highlighted in bold.  
\begin{table}[htbp]
\caption{Total Computation Time (s) of full grid SRP-PHAT, CFRC, BP, and MC Methods Under Different Levels for 3751 Points in Simulation. Bold means better.}

\begin{center}
\setlength{\tabcolsep}{4pt}
\renewcommand{\arraystretch}{1.2} 
\begin{tabular}{|c|c|c|c|c|} 
\hline
\textbf{Level} & \textbf{SRP-PHAT \cite{Knapp1976GCC}} & \textbf{CFRC \cite{Do2007cfrc}} & \textbf{BP (ASAP)} & \textbf{MC (ASAP)} \\
\hline
level 3 & 172.29 & 48.32 & \textbf{44.53} & 45.33 \\
\hline
level 4 & 723.31 & 61.19 & \textbf{53.96} & 56.18 \\
\hline
level 5 & 3987.86 & 92.81 & \textbf{73.31} & 76.53 \\
\hline
\end{tabular}
\label{tab:time_comparison}
\end{center}
\end{table}
\vspace{-8pt} 
 Table \ref{tab:time_comparison} shows that for the total computation task of 3751 points under different levels in the simulation, CFRC, ASAP (with BP or MC) exhibit favorable computational efficiency compared with  full grid SRP-PHAT. BP achieves the lowest computation time across all levels, while the computation time of full grid SRP-PHAT increases drastically as the level rises. Collectively, these results indicate that the proposed methods outperform the baseline methods in both computational efficiency and angular estimation accuracy, thereby verifying the effectiveness of the two-stage strategy in balancing multiple performance metrics.
 
\section{REAL-WORLD EXPERIMENTS}
\label{sec:real}
\begin{figure}[t]
  \centering
  \includegraphics[width=0.65\columnwidth]{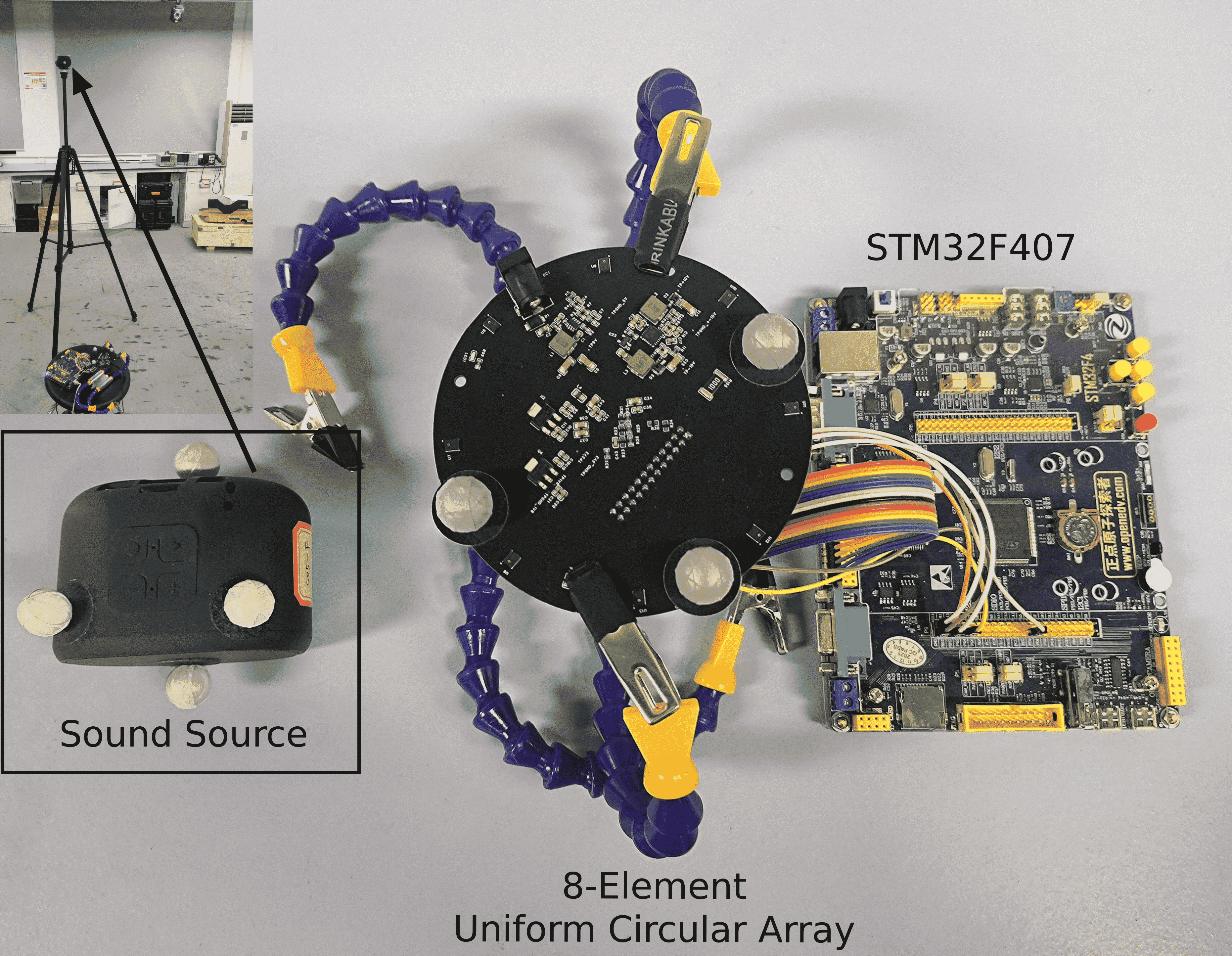}
  \caption{Experimental platform setup. An 8-microphone uniform circular array is used for sound source localization. The experimental environment and sound source are also illustrated.}
  \label{fig:platform_setup}
\end{figure}

To validate the practical performance of the proposed method, a series of real-world experiments were conducted in a typical office environment (approximately $4 \times 5 \times 3$ m, with reverberation time $T_{60} \approx 0.5$ s, ambient noise below 30 dB). We used a custom-designed 8-element UCA with a diameter of 88.8 mm placed on the horizontal plane (see Fig.~\ref{fig:platform_setup}). To evaluate the performance of different algorithms under consistent conditions, the array was positioned 0.5~m above the ground, and the sound source placed on the tripod was centered on the array at a distance of 1 to 2 meters and varied in azimuth and elevation angles. The array and source positions were tracked by a motion capture system and synchronized with the recorded audio. 

The duration of each audio recording of human voices is 2 seconds, with a sampling frequency of 50~kHz. 
Speech samples from 5 different speakers (3 males and 2 females) were used to enhance signal diversity. Table~\ref{tab:mae_results2} summarizes the localization errors of the proposed algorithm compared with baseline methods. Table \ref{tab:time_comparison2} summarizes total computation time of full grid SRP-PHAT, CFRC, ASAP (with BP or MC) under different levels for 523 recording data.

\begin{table}[htbp]
\caption{RMSE (deg) of Full Grid SRP-PHAT, CFRC, and ASAP Methods for 523 Recording Data. Bold means better.}
\begin{center}
\setlength{\tabcolsep}{4pt}
\renewcommand{\arraystretch}{1.3}
\begin{tabular}{|c|c|c|c|c|}
\hline
\textbf{Level} & \textbf{SRP-PHAT \cite{Knapp1976GCC}} & \textbf{CFRC \cite{Do2007cfrc}} & \textbf{BP (ASAP)} & \textbf{MC (ASAP)} \\
\hline
level 5 & 8.90 & 9.23 & \textbf{8.83} & 9.20 \\
\hline
\end{tabular}
\label{tab:mae_results2}
\end{center}
\end{table}
\vspace{-4pt} 

\begin{table}[htbp]
\caption{Total Computation Time (s) of Full Grid SRP-PHAT, CFRC, and ASAP for 523 Recording Data. Bold means better.}
\begin{center}
\setlength{\tabcolsep}{4pt}
\renewcommand{\arraystretch}{1.2}
\begin{tabular}{|c|c|c|c|c|}
\hline
\textbf{Level} & \textbf{SRP-PHAT \cite{Knapp1976GCC}} & \textbf{CFRC \cite{Do2007cfrc}} & \textbf{BP (ASAP)} & \textbf{MC (ASAP)} \\
\hline
level 3 & 67.38 & 18.91 & \textbf{17.28} & 17.73 \\
\hline
level 4 & 282.42 & 23.94 & \textbf{21.02} & 21.91 \\
\hline
level 5 & 1556.55 & 36.23 & \textbf{28.58} & 29.82 \\
\hline
\end{tabular}
\label{tab:time_comparison2}
\end{center}
\end{table}
\vspace{-8pt} 

Tables \ref{tab:mae_results2} and \ref{tab:time_comparison2} indicate that for 523 recordings at level 5, the proposed methods outperform baselines in both localization accuracy and computational efficiency. In terms of RMSE, our BP algorithm achieves the lowest value of 8.83°, surpassing SRP-PHAT (8.90°), CFRC (9.23°), and even the MC variant (9.20°). These results further validate the performance and robustness of ASAP in real reverberant environments. Table \ref{tab:time_comparison2} presents computation times for the 523 recordings across levels 3, 4, and 5. SRP-PHAT is significantly less efficient across all levels, with its time at level 5 reaching 1555.91 seconds while CFRC, BP, and MC only take 36.23, 28.58, and 29.82 seconds respectively. BP achieves the shortest computation time at each level (bolded). These real-world results align with simulations, confirming ASAP’s improvements in accuracy and efficiency.



\section{CONCLUSIONS}
\label{sec:concl}



This paper presents ASAP, an efficient approach to planar microphone array DOA estimation in 3D. In the first stage, ASAP performs a CFRC-style search within azimuthal strips to lock the azimuth angles while retaining multiple maxima with spherical caps. In the second stage, it refines elevation along the great-circle arc between two close candidates. By doing so, ASAP improves computational efficiency without sacrificing DOA estimation accuracy. Extensive simulations and experiments have been conducted to verify the efficacy and merits of the proposed method with respect to existing methods. In simulations, ASAP reduced runtime by an average of 13.56\% and RMSE by 5.87\%, compared to CFRC. For real experiments using 523 speech recordings with an 8-microphone UCA, ASAP decreased runtime by 13.98\% and RMSE by 4.33\%, compared to CFRC. More importantly, in both simulations and experiments, ASAP's performance is comparable to that of SRP-PHAT. 

While the performance gains over CFRC may appear incremental, ASAP can guarantee both accuracy and efficiency, which is critical for computing resource-constrained systems or far-field scenarios. For example, in the latter case, even marginal improvements in elevation estimation can help prevent significant accumulation errors that often lead to target tracking failures. Moreover, in its current form, ASAP relies on the trial-and-error tuning of a few key parameters, such as strip widths and subdivision levels. Our future research will explore how to integrate lightweight deep learning models with ASAP, aiming for a more robust and systematic approach. Additionally, we plan to extend ASAP to  complex scenarios involving multiple and/or moving sound sources.


\vspace{12pt}

\end{document}